\documentclass[11p]{article}

\usepackage{amsfonts}
\usepackage{amsmath}
\usepackage{amscd}
\usepackage{epsf}
\usepackage{times}
\usepackage{latexsym}
\usepackage{graphics}
\usepackage{graphicx}

\begin{document}
\title{Travelling wave solutions of BBM-like equations
by means of factorization}
\date{}
\maketitle
\begin{center}
\c{S}. Kuru\\
{\it Department of Physics, Faculty of Science, Ankara University,
06100 Ankara, Turkey}
\end{center}

\begin{abstract}
In this work, we apply the factorization technique to the
Benjamin-Bona-Mahony like equations in order to get travelling wave
solutions. We will focus on some special cases for which $m\neq n$,
and we will obtain these solutions in terms of Weierstrass
functions.

Email: kuru@science.ankara.edu.tr

\end{abstract}


\section{Introduction}

In this paper, we will consider the Benjamin-Bona-Mahony (BBM)
\cite{benjamin} like equation ($B(m,n)$) with a fully nonlinear
dispersive term of the form
\begin{equation}
u_{t}+u_{x}+a\,(u^m)_x-(u^n)_{xxt}=0, \quad\quad m,\,n>1,\,\,m \neq
n\, .\label{1.3}
\end{equation}
This equation is similar to the nonlinear dispersive equation
$K(m,n)$,
\begin{equation}
u_{t}+(u^m)_x+(u^n)_{xxx}=0, \quad\quad m>0,\,1<n\leq3 \label{1.2}
\end{equation}
which has been studied in detail by P. Rosenau and J.M. Hyman
\cite{rosenau}. In the literature there are many studies dealing
with the travelling wave solutions of the $K(m,n)$ and $B(m,n)$
equations, but in general they are restricted to the case $m=n$
\cite{rosenau,rosenau1,wazwaz,wazwaz1,wazwaz2,wazwaz3,wazwaz5,taha,ludu,wazwaz4,yadong,wang,kuru}.
When $m\neq n$, the solutions of $K(m,n)$ were investigated in
\cite{rosenau,rosenau1}. Our aim here is just to search for
solutions of the equations $B(m,n)$, with $m\neq n$, by means of the
factorization method.

We remark that this method \cite{pilar1,pilar,Perez,pilar2}, when it
is applicable, allows to get directly and systematically a wide set
of solutions, compared with other methods used  in the BBM
equations. For example, the direct integral method used by C. Liu
\cite{liu} can only be applied to the $B(2,1)$ equation. However,
the factorization technique can be used to more equations than the
direct integral method and also, in some cases, it gives rise to
more general solutions than the sine-cosine and the tanh methods
\cite{wazwaz4,yadong,wazwaz6}. This factorization approach to find
travelling wave solutions of nonlinear equations has been extended
to third order nonlinear ordinary differential equations (ODE's) by
D-S. Wang and H. Li \cite{li}.

When we look for the travelling wave solutions of Eq.~(\ref{1.3}),
first we reduce the form of the $B(m,n)$ equation to a second order
nonlinear ODE and then, we can immediately apply factorization
technique. Here, we will assume $m \neq n$, since the case $m = n$
has already been examined in a previous article following this
method \cite{kuru}.

This paper is organized as follows. In section 2 we introduce
factorization technique for a special type of the second order
nonlinear ODE's. Then, we apply straightforwardly the factorization
to the related second order nonlinear ODE to get travelling wave
solutions of $B(m,n)$ equation in section 3. We obtain the solutions
for these nonlinear ODE's and the $B(m,n)$ equation in terms of
Weierstrass functions in section 4. Finally, in section 5 we will
add some remarks.

\section{Factorization of nonlinear second
order ODE's}

Let us consider, the following nonlinear second order ODE
\begin{equation}\label{9}
 \frac{d^2 W}{d \theta^2}-\beta \frac{d W}{d \theta}+F(W)=0
\end{equation}
where $\beta$ is constant and $F(W)$ is an arbitrary function of
$W$. The factorized form of this equation can be written as
\begin{equation}\label{10}
\left[\frac{d}{d \theta}-f_2(W,\theta)\right]\left[\frac{d}{d
\theta}-f_1(W,\theta)\right] W(\theta)=0\,.
\end{equation}
Here, $f_1$ and $f_2$ are unknown functions that may depend
explicitly on $W$ and $\theta$. Expanding (\ref{10}) and comparing
with (\ref{9}), we obtain the following consistency conditions
\begin{equation}\label{12}
f_1f_2=\frac{F(W)}{W}+\frac{\partial f_1}{\partial \theta}, \qquad
f_2+\frac{\partial(W f_1)}{\partial W}=\beta.
\end{equation}
If we solve (\ref{12}) for $f_{1}$ or $f_{2}$, it will supply us to
write a compatible first order ODE
\begin{equation}\label{14}
\left[\frac{d}{d \theta}-f_1(W,\theta)\right] W(\theta)=0
\end{equation}
that provides a solution for the nonlinear ODE (\ref{9})
\cite{pilar1,pilar,Perez,pilar2}. In the applications of this paper
$f_{1}$ and $f_{2}$ will depend only on $W$.

\section{Factorization of the BBM-like equations
}

When Eq. (\ref{1.3}) has the travelling wave solutions in the form
\begin{equation}\label{15}
u(x,t)=\phi(\xi),\quad\quad \xi=hx+wt
\end{equation}
where $h$ and $w$ are real constants, substituting (\ref{15}) into
(\ref{1.3}) and after integrating, we get the reduced form of Eq.
(\ref{1.3}) to the second order nonlinear ODE
\begin{equation}\label{16}
(\phi^n)_{\xi\xi}-A\,\phi-B\,\phi^m+D=0\,.
\end{equation}
Notice that the constants in Eq. (\ref{16}) are
\begin{equation}\label{17}
A=\frac{h+w}{h^2\,w},\quad\quad B=\frac{a}{h\,w},\quad\quad
D=\frac{R}{h^2\,w}
\end{equation}
and $R$ is an integration constant. Now, if we introduce the
following natural transformation of the dependent variable
\begin{equation}\label{18}
\phi^n(\xi)=W(\theta),\quad\quad\xi=\theta
\end{equation}
Eq.~(\ref{16}) becomes
\begin{equation}\label{19}
\frac{d^2 W}{d \theta^2}-A\,W^{\frac{1}{n}}-B\,W^{\frac{m}{n}}+D=0.
\end{equation}
Now, we can apply the factorization technique to Eq.~(\ref{19}).
Comparing Eq.~(\ref{9}) and Eq.~(\ref{19}), we have $\beta=0$ and
\begin{equation}\label{20}
F(W)=-(A\,W^{\frac{1}{n}}+B\,W^{\frac{m}{n}}-D)\,.
\end{equation}
Then, from (\ref{12}) we get only one consistency condition
\begin{equation}\label{23}
f_1^2+f_1\,W\frac{df_1}{dW}-A\,W^{\frac{1-n}{n}}-B\,
W^{\frac{m-n}{n}}+D\,W^{-1}=0\,
\end{equation}
whose solutions are
\begin{equation}\label{24}
f_1(W)=\pm\frac{1}{W}\sqrt{\frac{2\,n\,A}{n+1}\,W^{\frac{n+1}{n}}+
\frac{2\,n\,B}{m+n}\,W^{\frac{m+n}{n}}-2\, D\,W+C}\,
\end{equation}
where $C$ is an integration constant. Thus, the first order ODE
(\ref{14}) takes the form
\begin{equation}\label{25}
\frac{dW}{d
\theta}\mp\sqrt{\frac{2\,n\,A}{n+1}\,W^{\frac{n+1}{n}}+\frac{2\,n\,B}{m+n}\,W^{\frac{m+n}{n}}-2\,
D\,W+C}=0\,.
\end{equation}
In order to solve this equation for $W$ in a more general way, let
us take $W$ in the form $W=\varphi^p,\,p\neq0,1$, then, the first
order ODE (\ref{25}) is rewritten in terms of $\varphi$ as
\begin{equation}\label{26}
(\frac{d\varphi}{d \theta})^2=
\frac{2\,n\,A}{p^2\,(n+1)}\,\varphi^{p(\frac{1-n}{n})+2}+
\frac{2\,n\,B}{p^2\,(m+n)}\,\varphi^{p(\frac{m-n}{n})+2}-\frac{2\,D}{p^2}\,\varphi^{2-p}
+\frac{C}{p^2}\,\varphi^{2-2\,p}\,.
\end{equation}
If we want to guarantee the integrability of (\ref{26}), the powers
of $\varphi$ have to be integer numbers between $0$ and $4$
 \cite{ince}. Having
in mind the conditions on $n, m$ ($n\neq m >1$) and $p$ ($p\neq0$),
we have the following possible cases:
\begin{itemize}
\item If $C=0,\,\,D=0$, we can choose $p$ and $m$ in the following
way
\begin{equation}\label{27a}
p=\pm \frac{2n}{1-n} \quad {\rm{with}} \quad
m=\frac{n+1}{2},\frac{3\,n-1}{2}, 2\,n-1
\end{equation}
and
\begin{equation}\label{27b}
p=\pm \frac{n}{1-n}\quad {\rm{with}} \quad m= 2\,n-1,3\,n-2\,.
\end{equation}

It can be checked that the two choices of sign in (\ref{27a}) and
(\ref{27b}) give rise to the same solutions for Eq.~(\ref{1.3}).
Therefore, we will consider only one of them. Then, taking $p=-
\frac{2n}{1-n}$, Eq. (\ref{26}) becomes
\begin{equation}\label{29}
(\frac{d\varphi}{d
\theta})^2=\frac{A\,(n-1)^2}{2\,n\,(n+1)}+\frac{B\,(n-1)^2}{n\,(3\,n+1)}\,\varphi,
\quad m=\frac{n+1}{2}
\end{equation}
\begin{equation}\label{30}
(\frac{d\varphi}{d
\theta})^2=\frac{A\,(n-1)^2}{2\,n\,(n+1)}+\frac{B\,(n-1)^2}{n\,(5\,n-1)}\,\varphi^3,
\quad m=\frac{3\,n-1}{2}
\end{equation}
\begin{equation}\label{31}
(\frac{d\varphi}{d
\theta})^2=\frac{A\,(n-1)^2}{2\,n\,(n+1)}+\frac{B\,(n-1)^2}{n\,(3\,n-1)}\,\varphi^4,
\quad m=2\,n-1
\end{equation}
and for $p=- \frac{n}{1-n}\,$,
\begin{equation}\label{32}
(\frac{d\varphi}{d
\theta})^2=\frac{2\,A\,(n-1)^2}{n\,(n+1)}\,\varphi+\frac{2\,B\,(n-1)^2}{n\,(3\,n-1)}\,\varphi^3,
\quad m=2\,n-1
\end{equation}
\begin{equation}\label{33}
(\frac{d\varphi}{d
\theta})^2=\frac{2\,A\,(n-1)^2}{n\,(n+1)}\,\varphi+\frac{B\,(n-1)^2}{n\,(2\,n-1)}\,\varphi^4,
\quad m=3\,n-2\,.
\end{equation}
\item If $C=0$, we have the special cases,
$p=\pm 2$, $n=2$ with $m=3,4$.

Due to the same reason in the above case, we will consider only
$p=2$. Then, Eq.~(\ref{26}) takes the form:
\begin{equation}\label{e23}
(\frac{d\varphi}{d
\theta})^2=-\frac{D}{2}+\frac{A}{3}\,\varphi+\frac{B}{5}\,\varphi^3,
\quad m=3
\end{equation}
\begin{equation}\label{e24}
(\frac{d\varphi}{d
\theta})^2=-\frac{D}{2}+\frac{A}{3}\,\varphi+\frac{B}{6}\,\varphi^4,
\quad m=4\,.
\end{equation}

\item If $A=C=0$, we have $p=\pm 2$ with
$m=\displaystyle \frac{n}{2},\frac{3\,n}{2},2\,n$.

In this case,  for $p=2$, Eq.~(\ref{26}) has the following form:
\begin{equation}\label{e2n2}
(\frac{d\varphi}{d
\theta})^2=-\frac{D}{2}\,\varphi^4+\frac{B}{3}\,\varphi^3, \quad
m=\frac{n}{2}
\end{equation}
\begin{equation}\label{e3n2}
(\frac{d\varphi}{d
\theta})^2=-\frac{D}{2}\,\varphi^4+\frac{B}{5}\,\varphi, \quad
m=\frac{3\,n}{2}
\end{equation}
\begin{equation}\label{e22n}
(\frac{d\varphi}{d \theta})^2=-\frac{D}{2}\,\varphi^4+\frac{B}{6},
\quad m=2\,n\,.
\end{equation}

\item If $A=0$, we have $p=\pm 1$ with $m=2\,n,3\,n$.

Here, also we will take only the case $p=1$, then, we will have the
equations:
\begin{equation}\label{e2n}
(\frac{d\varphi}{d
\theta})^2=-2\,D\,\varphi+\frac{2}{3}\,B\,\varphi^3+C\varphi^4,
\quad m=2n
\end{equation}
\begin{equation}\label{e3n}
(\frac{d\varphi}{d
\theta})^2=-2\,D\,\varphi+\frac{B}{2}\,\varphi^4+C, \quad m=3n\,.
\end{equation}
\item If $A=D=0$, we have $p=\displaystyle \pm \frac{1}{2}$ with $m=3\,n,5\,n$.

Thus, for $p=\displaystyle \frac{1}{2}$, Eq.~(\ref{26}) becomes:
\begin{equation}\label{e3n1}
(\frac{d\varphi}{d \theta})^2=2\,B\,\varphi^3+4\,C\varphi, \quad
m=3n
\end{equation}
\begin{equation}\label{e3nn}
(\frac{d\varphi}{d
\theta})^2=\frac{4}{3}\,B\,\varphi^4+4\,C\,\varphi, \quad m=5n\,.
\end{equation}

\end{itemize}
\section{Travelling wave solutions for BBM-like equations}

In this section, we will obtain the solutions of the differential
equations (\ref{29})-(\ref{33}) in terms of Weierstrass function,
$\wp(\theta;g_{2},g_{3})$, which allow us  to get the travelling
wave solutions of $B(m,n)$ equations (\ref{1.3}). The rest of
equations (\ref{e23})-(\ref{e3nn}) can be dealt with a similar way,
but they will not be worked out here for the sake of shortness.

First, we will give some properties of the $\wp$ function which will
be useful in the following \cite{Bateman,watson}.

\subsection{Relevant properties of the $\wp$ function}
Let us consider a differential equation with a quartic polynomial
\begin{equation}\label{ef}
\big(\frac{d\varphi}{d\theta}\big)^2 =P(\varphi) =
a_{0}\,\varphi^4+4\,a_{1}\,\varphi^3+6\,a_{2}\,\varphi^2+4\,a_{3}\,\varphi+a_{4}\,.
\end{equation}
The solution of this equation can be written in terms of the
Weierstrass function  where the invariants $g_2$ and $g_3$ of
(\ref{ef}) are
\begin{equation}\label{gg}
g_{2}= a_{0}\,a_{4}-4\,a_{1}\,a_{3}+3\,a_{2}^2,\ \ g_{3}=
a_{0}\,a_{2}\,a_{4}+2\,a_{1}\,a_{2}\,a_{3}-a_{2}^{3}-a_{0}\,a_{3}^2-a_{1}^{2}\,a_{4}
\end{equation}
and the discriminant is given by $\Delta=g_2^3-27\,g_3^2$. Then, the
solution $\varphi$ can be found as
\begin{equation}\label{x}
\varphi(\theta)=\varphi_0+\frac{1}{4}P_\varphi(\varphi_0)\left(\wp(\theta;g_{2},g_{3})-
\frac{1}{24}P_{\varphi\varphi}(\varphi_0)\right)^{-1}
\end{equation}
where the subindex in $P_{\varphi}(\varphi_0)$ denotes the
derivative with respect to $\varphi$, and $\varphi_0$ is one of the
roots of the polynomial $P(\varphi)$ (\ref{ef}). Depending of the
selected root $\varphi_0$, we will have a solution with a different
behavior \cite{kuru}.

Here, also we want to recall some other properties of the
Weierstrass functions
\cite{stegun}:

i) The case $g_2=1$ and $g_3=0$ is called lemniscatic case
\begin{equation}\label{lc}
\wp(\theta;g_{2},0)=g_2^{1/2}\,\wp(\theta\,g_{2}^{1/4};1,0),\qquad
g_2>0\,
\end{equation}

ii) The case $g_2=-1$ and $g_3=0$ is called pseudo-lemniscatic case
\begin{equation}\label{plc}
\wp(\theta;g_{2},0)=|g_2|^{1/2}\,\wp(\theta\,|g_{2}|^{1/4};-1,0),\qquad
g_2<0\,
\end{equation}

iii) The case $g_2=0$ and $g_3=1$ is called equianharmonic case
\begin{equation}\label{ec}
\wp(\theta;g_{2},0)=g_3^{1/3}\,\wp(\theta\,g_{3}^{1/6};0,1),\qquad
g_3>0\,.
\end{equation}

Once obtained the solution $W(\theta)$, taking into account
(\ref{15}), (\ref{18}) and $W=\varphi^{p}$, the solution of
Eq.~(\ref{1.3}) is obtanied as
\begin{equation}\label{uxt}
u(x,t)=\phi(\xi)=W^{\frac{1}{n}}(\theta)=\varphi^{\frac{p}{n}}(\theta),\quad\quad\theta=\xi=h\,x+w\,t.
\end{equation}

\subsection{The case $C=0,\,D=0$, $\displaystyle p=-\frac{2n}{1-n}$}
\begin{itemize}
\item  $m=\displaystyle \frac{n+1}{2}$

\noindent Equation (\ref{29}) can be expressed as
\begin{equation}
(\frac{d\varphi}{d\theta})^2=P(\varphi)=\frac{A\,(n-1)^2}
{2\,n\,(n+1)}+\frac{B\,(n-1)^2}{n\,(3\,n+1)}\,\varphi
\end{equation}
and from $P(\varphi)=0$, we get the root of this polynomial
\begin{equation}\label{f01}
 \varphi_0=-\frac{A\,(3\,n+1)}{2\,B\,(n+1)}\,.
\end{equation}
The invariants (\ref{gg}) are: $g_{2}=g_{3}=0$, and $\Delta=0$.
Therefore, having in mind $\wp(\theta;0,0)=\displaystyle
\frac{1}{\theta^2}$, we can find the solution of (\ref{29}) from
(\ref{x}) for $\varphi_0$, given by (\ref{f01}),
\begin{equation}\label{35}
\varphi(\theta)=
\frac{B^2\,(n-1)^2\,(n+1)\,\theta^2-2\,A\,n\,(3\,n+1)^2}{4\,B\,n\,(n+1)\,(3\,n+1)}\,.
\end{equation}
Now, the solution of Eq.~(\ref{1.3}) reads from (\ref{uxt})
\begin{equation}\label{u1}
u(x,t)=\left[\frac{B^2\,(n-1)^2\,(n+1)\,(h\,x+w\,t)^2-2\,A\,n\,(3\,n+1)^2}
{4\,B\,n\,(n+1)\,(3\,n+1)}\right]^{\frac{2}{n-1}}\,.
\end{equation}
\item  $m=\displaystyle \frac{3\,n-1}{2}$

\noindent In this case, our equation to solve is (\ref{30}) and the
polynomial has the form
\begin{equation}
P(\varphi)=\frac{A\,(n-1)^2}{2\,n\,(n+1)}+\frac{B\,(n-1)^2}{n\,(5\,n-1)}\,\varphi^3
\end{equation}
with one real root:
$\varphi_0=\left(\frac{-A\,(5\,n-1)}{2\,B\,(n+1)}\right)^{1/3}$.
Here, the discriminant is different from zero with the invariants
\begin{equation}
g_2=0,\qquad
g_3=\frac{-A\,B^2\,(n-1)^6}{32\,n^3\,(n+1)\,(5\,n-1)^2}\,.
\end{equation}
Then, the solution of (\ref{30}) is obtained by (\ref{x}) for
$\varphi_0$,
\begin{equation}\label{phi2}
\varphi=\varphi_0\,\left[\frac{4\,n\,(5\,n-1)\,\wp(\theta;0,g_3)+2\,B\,(n-1)^2\,\varphi_0
}{4\,n\,(5\,n-1)\,\wp(\theta;0,g_3)-B\,(n-1)^2\,\varphi_0}\right]
\end{equation}
and we get the solution of Eq.~(\ref{1.3}) from (\ref{uxt}) as
\begin{equation}\label{u2}
u(x,t)=\left[\varphi_0^2\,\left(\frac{4\,n\,(5\,n-1)\,\wp(h\,x+w\,t;0,g_3)+2\,B\,(n-1)^2\,\varphi_0
}{4\,n\,(5\,n-1)\,\wp(h\,x+w\,t;0,g_3)-B\,(n-1)^2\,\varphi_0}\right)^2\right]^{\frac{1}{n-1}}
\end{equation}
with the conditions: $A<0,g_3>0$, for
$\varphi_0=\left(\frac{-A\,(5\,n-1)}{2\,B\,(n+1)}\right)^{1/3}$.
Using the relation (\ref{ec}), we can write the solution (\ref{u2})
in terms of equianharmonic case of the Weierstrass function:
\begin{equation}\label{u21}
u(x,t)=\left[\left(\frac{-A\,(5\,n-1)}{2\,B\,(n+1)}\right)^{2/3}\,
\left(\frac{2^{2/3}\,\wp((h\,x+w\,t)\,g_3^{1/6};0,1)+2
}{2^{2/3}\,\wp((h\,x+w\,t)\,g_3^{1/6};0,1)-1}\right)^2\right]^{\frac{1}{n-1}}\,.
\end{equation}

\item  $m=2\,n-1$

\noindent In Eq.~(\ref{31}), the quartic polynomial is
\begin{equation}
P(\varphi)=\frac{A\,(n-1)^2}{2\,n\,(n+1)}+\frac{B\,(n-1)^2}{2\,n\,(3\,n-1)}\,\varphi^4
\end{equation}
and has two real roots:
$\varphi_0=\pm\left(\frac{-A\,(3\,n-1)}{B\,(n+1)}\right)^{1/4}$ for
$A<0,\,B>0$ or $A>0,\,B<0$. In this case, the invariants are
\begin{equation}
g_2=\frac{A\,B\,(n-1)^4}{4\,n^2\,(n+1)\,(3\,n-1)},\qquad g_3=0\,.
\end{equation}
Here, also the discriminant is different from zero, $\Delta\neq0$.
We obtain the solution of (\ref{31}) from (\ref{x}) for $\varphi_0$,
\begin{equation}\label{phi3}
\varphi=\varphi_0\,\left[\frac{4\,n\,(n+1)\,\varphi_0^2\,\wp(\theta;g_2,0)-A\,(n-1)^2
}{4\,n\,(n+1)\,\varphi_0^2\,\wp(\theta;g_2,0)+A\,(n-1)^2}\right]
\end{equation}
and we get the solution of Eq.~(\ref{1.3}) from (\ref{uxt}) as
\begin{equation}\label{u3}
u(x,t)=\left[\varphi_0^2\,\left(\frac{4\,n\,(n+1)\,\varphi_0^2\,\wp(h\,x+w\,t;g_2,0)-A\,(n-1)^2
}{4\,n\,(n+1)\,\varphi_0^2\,\wp(h\,x+w\,t;g_2,0)+A\,(n-1)^2}\right)^2\right]^{\frac{1}{n-1}}
\end{equation}
with the conditions for real solutions: $A<0,\,B>0,\,g_2<0$ or
$A>0,\,B<0,\,g_2<0$.

Having in mind the relation (\ref{plc}), the solution (\ref{u3}) can
be expressed in terms of the pseudo-lemniscatic case of the
Weierstrass function:
\begin{equation}\label{u32}
u(x,t)=\left[\left(\frac{-A\,(3\,n-1)}{B\,(n+1)}\right)^{1/2}\,
\left(\frac{2\,\wp((h\,x+w\,t)|g_2|^{1/4};-1,0)+1
}{2\,\wp((h\,x+w\,t)|g_2|^{1/4};-1,0)-1}\right)^2\right]^{\frac{1}{n-1}}
\end{equation}
for $A<0,\,B>0,\,g_2<0$ and
\begin{equation}\label{u31}
u(x,t)=\left[\left(\frac{-A\,(3\,n-1)}{B\,(n+1)}\right)^{1/2}\,
\left(\frac{2\,\wp((h\,x+w\,t)|g_2|^{1/4};-1,0)-1
}{2\,\wp((h\,x+w\,t)|g_2|^{1/4};-1,0)+1}\right)^2\right]^{\frac{1}{n-1}}
\end{equation}
for $A>0,\,B<0,\,g_2<0$.
\end{itemize}

\subsection{The case $C=0,\,D=0$, $\displaystyle p=- \frac{n}{1-n}$}
\begin{figure}[h]
  \centering
\includegraphics[width=0.4\textwidth]{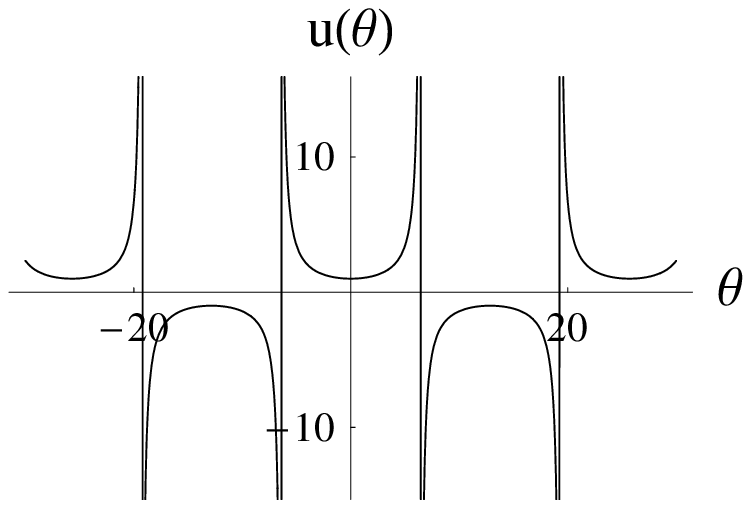}%
\hspace{1cm}%
  \includegraphics[width=0.4\textwidth]{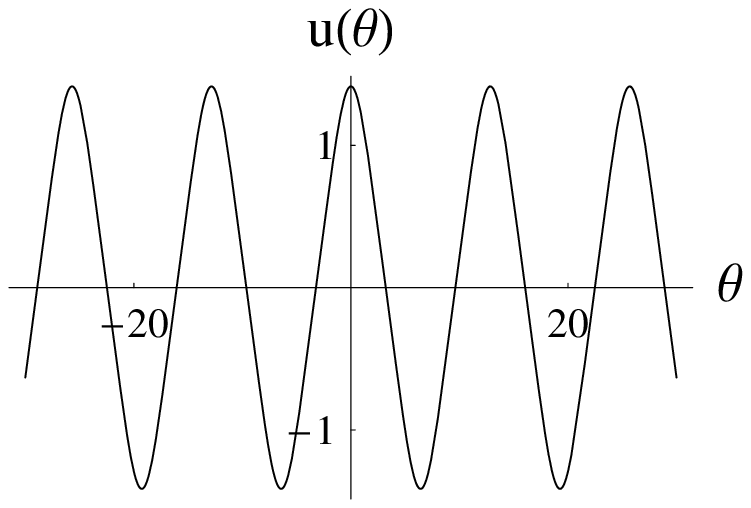}
\caption{The left figure corresponds to the solution (\ref{u31}) for
$h=-2$, $w=1$, $a=-1$, $n=3$, $m=5$ and the right one corresponds to
the solution (\ref{u32}) for $h=1$, $w=1$, $a=-1$, $n=3$, $m=5$.}
  \label{figuras1}
\end{figure}

\begin{figure}[h]
  \centering
\includegraphics[width=0.4\textwidth]{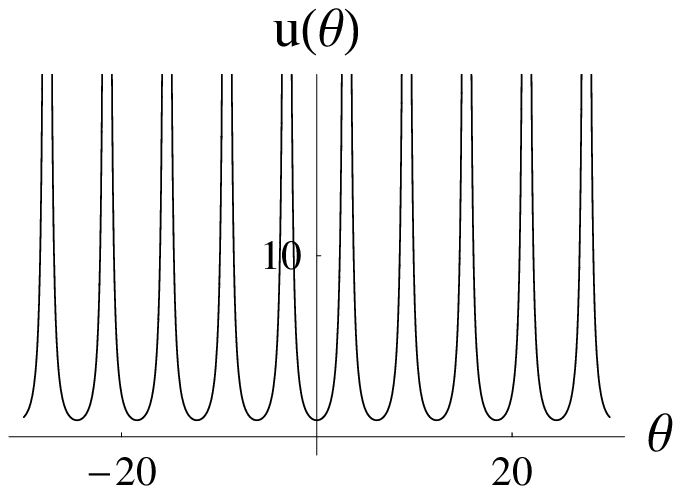}%
\hspace{1cm}%
  \includegraphics[width=0.4\textwidth]{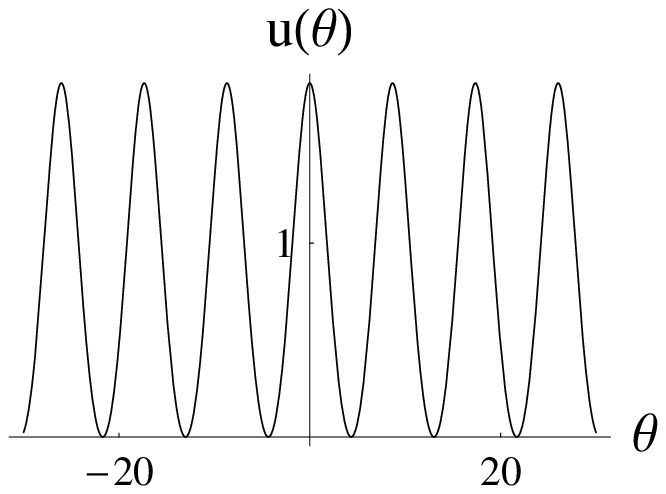}
\caption{The left figure corresponds to the solution (\ref{u31}) for
$h=-2$, $w=1$, $a=-1$, $n=2$, $m=3$ and the right one corresponds to
the solution (\ref{u32}) for $h=1$, $w=1$, $a=-1$, $n=2$, $m=3$.}
  \label{figuras111}
\end{figure}

\begin{figure}[h]
  \centering
\includegraphics[width=0.4\textwidth]{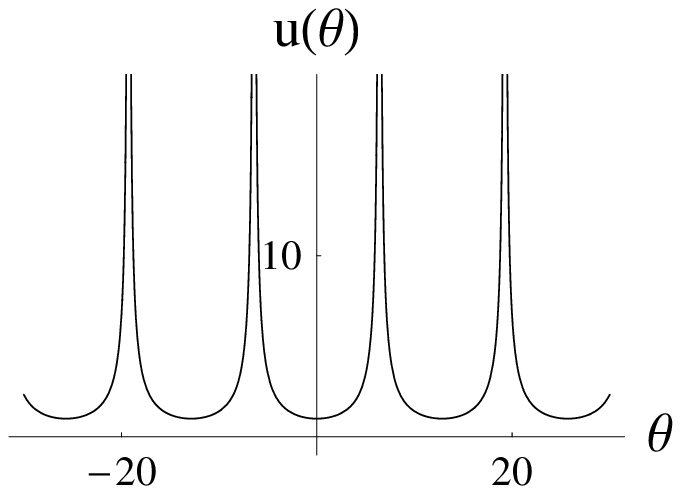}%
\hspace{1cm}%
  \includegraphics[width=0.4\textwidth]{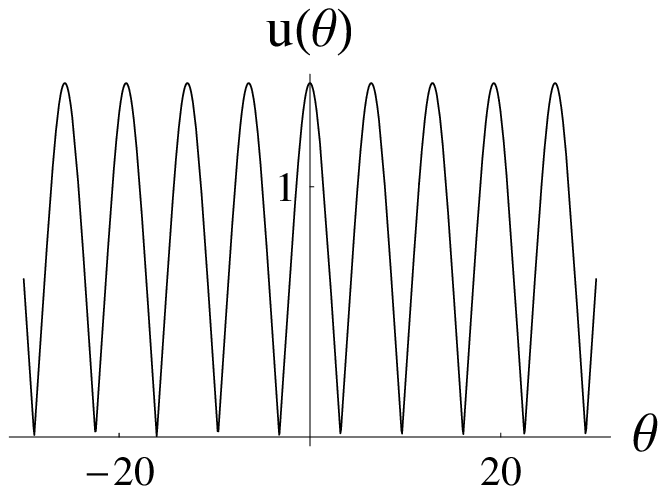}
\caption{The left figure corresponds to the solution (\ref{u41}) for
$h=-2$, $w=1$, $a=-1$, $n=3$, $m=5$  and the right one corresponds
to the solution (\ref{u42}) for $h=1$, $w=1$, $a=-1$, $n=3$, $m=5$.
}
  \label{figuras2}
\end{figure}

\begin{figure}[h]
  \centering
\includegraphics[width=0.4\textwidth]{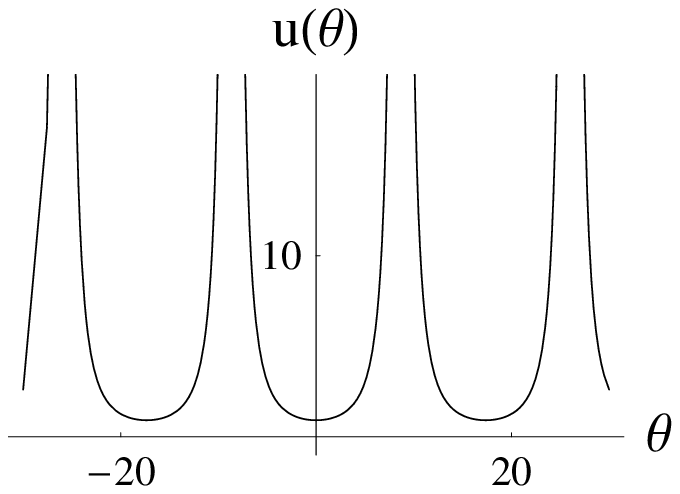}%
\hspace{1cm}%
  \includegraphics[width=0.4\textwidth]{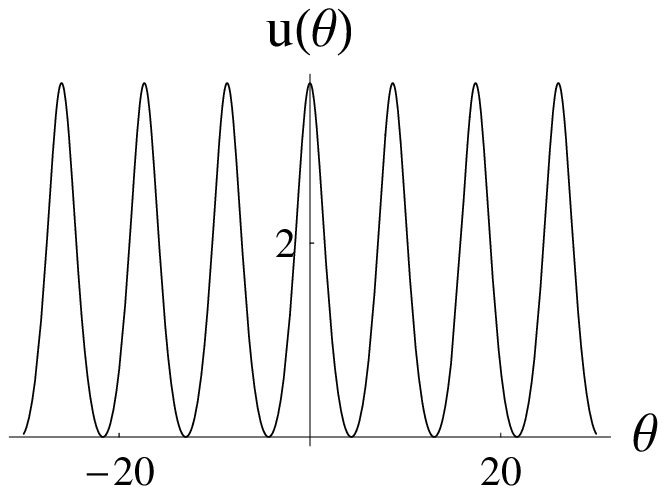}
\caption{The left figure corresponds to the solution (\ref{u41}) for
$h=-2$, $w=1$, $a=-1$, $n=2$, $m=3$  and the right one corresponds
to the solution (\ref{u42}) for $h=1$, $w=1$, $a=-1$, $n=2$, $m=3$.
}
  \label{figuras222}
\end{figure}

\begin{figure}[h]
  \centering
\includegraphics[width=0.4\textwidth]{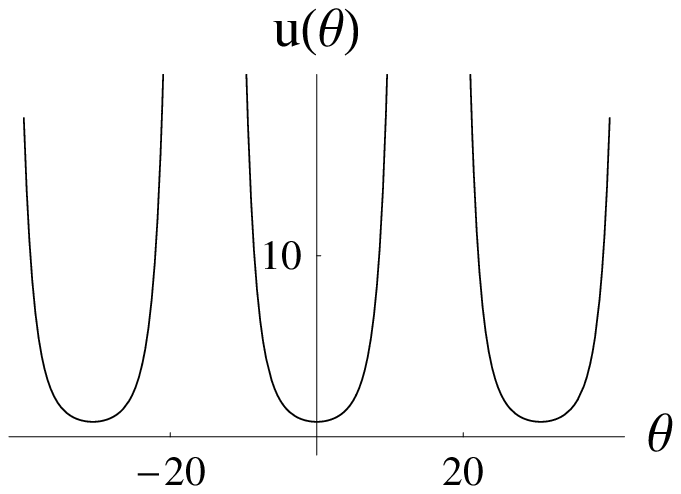}%
\hspace{1cm}%
  \includegraphics[width=0.4\textwidth]{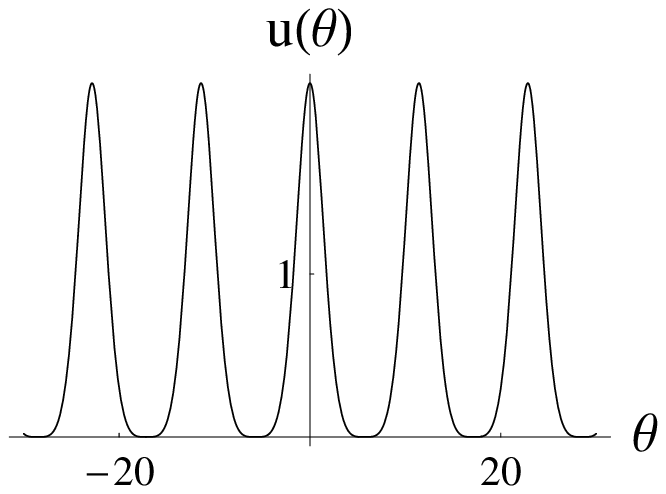}
\caption{The left figure corresponds to the solution  (\ref{u21})
for $h=-2$, $w=1$, $a=-1$, $n=2$, $m=5/2$ and the right one
corresponds to the solution (\ref{u51}) for $h=1$, $w=1$, $a=-1$,
$n=3/2$, $m=5/2$.}
  \label{figuras333}
\end{figure}

\begin{itemize}

\item  $m=2\,n-1$

\noindent Now, the polynomial is cubic
\begin{equation}
P(\varphi)=\frac{2\,A\,(n-1)^2}{n\,(n+1)}\,\varphi+\frac{2\,B\,(n-1)^2}{n\,(3\,n-1)}\,\varphi^3
\end{equation}
and has three distinct real roots: $\varphi_0=0$ and
$\varphi_0=\pm\left(\frac{-A\,(3\,n-1)}{B\,(n+1)}\right)^{1/2}$ for
$A<0,\,B>0$ or $A>0,\,B<0$. Now, the invariants are
\begin{equation}
g_2=\frac{-A\,B\,(n-1)^4}{n^2\,(n+1)\,(3\,n-1)},\qquad g_3=0
\end{equation}
and $\Delta\neq0$. The solution of (\ref{32}) is obtained from
(\ref{x}) for $\varphi_0$,
\begin{equation}\label{phi4}
\varphi=\varphi_0\,\left[\frac{2\,n\,(n+1)\,\varphi_0\,\wp(\theta;g_2,0)-A\,(n-1)^2
}{2\,n\,(n+1)\,\varphi_0\,\wp(\theta;g_2,0)+A\,(n-1)^2}\right]
\end{equation}
and substituting  (\ref{phi4}) in  (\ref{uxt}), we get the solution
of Eq.~(\ref{1.3}) as
\begin{equation}\label{u4}
u(x,t)=\left[\varphi_0\,\left(\frac{2\,n\,(n+1)\,\varphi_0\,\wp(h\,x+w\,t;g_2,0)-A\,(n-1)^2
}{2\,n\,(n+1)\,\varphi_0\,\wp(h\,x+w\,t;g_2,0)+A\,(n-1)^2}\right)\right]^{\frac{1}{n-1}}
\end{equation}
with the conditions: $A<0,\,B>0,\,g_2>0$ and $A>0,\,B<0,\,g_2>0$ for
$\varphi_0=\left(\frac{-A\,(3\,n-1)}{B\,(n+1)}\right)^{1/2}$. While
the root $\varphi_0=0$ leads to the trivial solution, $u(x,t)=0$,
the other root
$\varphi_0=-\left(\frac{-A\,(3\,n-1)}{B\,(n+1)}\right)^{1/2}$ gives
rise to imaginary solutions.

Now, we can rewrite the solution (\ref{u4}) in terms of the
lemniscatic case of the Weierstrass function using the relation
(\ref{lc}) in (\ref{u4}):
\begin{equation}\label{u41}
u(x,t)=\left[\left(\frac{-A\,(3\,n-1)}{B\,(n+1)}\right)^{1/2}\,
\left(\frac{2\,\wp((h\,x+w\,t)\,g_2^{1/4};1,0)+1
}{2\,\wp((h\,x+w\,t)\,g_2^{1/4};1,0)-1}\right)\right]^{\frac{1}{n-1}}
\end{equation}
for $A<0,\,B>0,\,g_2>0$ and
\begin{equation}\label{u42}
u(x,t)=\left[\left(\frac{-A\,(3\,n-1)}{B\,(n+1)}\right)^{1/2}\,
\left(\frac{2\,\wp((h\,x+w\,t)\,g_2^{1/4};1,0)-1
}{2\,\wp((h\,x+w\,t)\,g_2^{1/4};1,0)+1}\right)\right]^{\frac{1}{n-1}}
\end{equation}
for $A>0,\,B<0,\,g_2>0$.

\item  $m=3\,n-2$

\noindent In this case, we have also a quartic polynomial
\begin{equation}
P(\varphi)=\frac{2\,A\,(n-1)^2}{n\,(n+1)}\,\varphi+\frac{B\,(n-1)^2}{n\,(2\,n-1)}\,\varphi^4
\, .
\end{equation}
It has two real roots: $\varphi_0=0$ and
$\varphi_0=\left(-\frac{2\,A\,(2\,n-1)}{B\,(n+1)}\right)^{1/3}$. For
the equation (\ref{33}), the invariants are
\begin{equation}
g_2=0,\qquad g_3=\frac{-A^2\,B\,(n-1)^6}{4\,n^3\,(n+1)^2\,(2\,n-1)}
\end{equation}
and $\Delta\neq0$. Now, the solution of (\ref{33}) reads from
(\ref{x}) for $\varphi_0$,
\begin{equation}\label{phi5}
\varphi=\varphi_0\,\left[\frac{2\,n\,(n+1)\,\varphi_0\,\wp(\theta;0,g_3)-A\,(n-1)^2
}{2\,n\,(n+1)\,\varphi_0\,\wp(\theta;0,g_3)+2\,A\,(n-1)^2}\right]\,.
\end{equation}
Then, the solution of Eq.~(\ref{1.3}) is from (\ref{uxt}) as
\begin{equation}\label{u5}
u(x,t)=\left[\varphi_0\,\left(\frac{2\,n\,(n+1)\,\varphi_0\,\wp(h\,x+w\,t;0,g_3)-A\,(n-1)^2
}{2\,n\,(n+1)\,\varphi_0\,\wp(h\,x+w\,t;0,g_3)+2\,A\,(n-1)^2}\right)\right]^{\frac{1}{n-1}}
\end{equation}
with the conditions: $B<0,\,g_3>0$. Taking into account the relation
(\ref{ec}), this solution also can be expressed in terms of the
equianharmonic case of the Weierstrass function:
\begin{equation}\label{u51}
u(x,t)=\left[\left(-\frac{2\,A\,(2\,n-1)}{B\,(n+1)}\right)^{1/3}\,
\left(\frac{2^{2/3}\,\wp((h\,x+w\,t)\,g_3^{1/6};0,1)-1
}{2^{2/3}\,\wp((h\,x+w\,t)\,g_3^{1/6};0,1)+2}\right)\right]^{\frac{1}{n-1}}\,.
\end{equation}
\end{itemize}
We have also plotted these solutions for some special values in
Figs. (\ref{figuras1})-(\ref{figuras333}). We can appreciate that
for the considered cases, except the parabolic case (\ref{35}), they
consist in periodic waves, some are singular while others are
regular. Their amplitude is governed by the non-vanishing constants
$A,B$ and their formulas are given in terms of the special forms
(\ref{lc})-(\ref{ec}) of the $\wp$ function.

\section{Lagrangian and Hamiltonian}

Since Eq. (\ref{19}) is a motion-type, we can write the
corresponding Lagrangian
\begin{equation}\label{lag}
L_W=\frac{1}{2}\,W_{\theta}^2+\frac{A\,n}{n+1}\,W^\frac{n+1}{n}+\frac{B\,n}{m+n}\,W^\frac{m+n}{n}-D\,W\,
\end{equation}
and, the Hamiltonian $H_W=W_{\theta} P_W -L_W$ reads
\begin{equation}H_W(W,P_W,\theta)=\frac{1}{2}\left[P_W^2-\left(\frac{2\,A\,n}{n+1}\,W^\frac{n+1}{n}+
\frac{2\,B\,n}{m+n}\,W^\frac{m+n}{n}-2\,D\,W\right)\right]
\label{hamil}\end{equation} where the canonical momentum is
\begin{equation}P_W=\frac{\partial L_W}{\partial W_\theta}=W_\theta.\label{mo}\end{equation}
The independent variable $\theta$ does not appear explicitly in
(\ref{hamil}), then  $H_W$ is a constant of motion,  $H_W=E$, with
\begin{equation}
E=\frac{1}{2}
\left[\left(\frac{dW}{d\theta}\right)^2-\left(\frac{2\,A\,n}{n+1}\,W^\frac{n+1}{n}+
\frac{2\,B\,n}{m+n}\,W^\frac{m+n}{n}-2\,D\,W\right)\right].\label{ee}
\end{equation}
Note that this equation also leads to the first order ODE (\ref{25})
with the identification $C=2\,E$.  Now, the energy $E$ can be
expressed as a product of two independent constant of motions
\begin{equation}
E=\frac{1}{2}\, I_+\,I_- \label{5.1}
\end{equation}
where
\begin{equation}
I_{\pm}(z)=\left(W_\theta\mp
\sqrt{\frac{2\,A\,n}{n+1}\,W^\frac{n+1}{n}+
\frac{2\,B\,n}{m+n}\,W^\frac{m+n}{n}-2\,D\,W}\,\right) \,e^{\pm
S(\theta)} \label{const}
\end{equation}
and the phase $S(\theta)$ is chosen in such a way that
$I_\pm(\theta)$  be constants of motion ($dI_\pm(\theta)/d\theta=0
$)
\begin{equation}
S(\theta)=\int
\frac{A\,W^{\frac{1}{n}}+B\,W^{\frac{m}{n}}-D}{\sqrt{\frac{2\,A\,n}{n+1}\,W^\frac{n+1}{n}+
\frac{2\,B\,n}{m+n}\,W^\frac{m+n}{n}-2\,D\,W}}\,d\theta.
\end{equation}
\section{Conclusions}

In this paper, we have applied the factorization technique to the
$B(m,n)$ equations in order to get travelling wave solutions. We
have considered some representative cases of the $B(m,n)$ equation
for $m\neq n$. By using this method, we obtained the travelling wave
solutions in a very compact form, where the constants appear as
modulating the amplitude, in terms of some special forms of the
Weierstrass elliptic function: lemniscatic, pseudo-lemniscatic and
equiaharmonic. Furthermore, these solutions are not only valid for
integer $m$ and $n$ but also non integer $m$ and $n$. The case $m=n$
for the $B(m,n)$ equations has been examined by means of the
factorization technique in a previous paper \cite{kuru} where the
compactons and kink-like solutions recovering all the solutions
previously reported have been constructed. Here, for $m\neq n$,
solutions with compact support can also be obtained following a
similar procedure. We note that, this method is systematic and gives
rise to a variety of solutions for nonlinear equations. We have also
built the Lagrangian and Hamiltonian for the second order nonlinear
ODE corresponding to the travelling wave reduction of the $B(m,n)$
equation. Since the Hamiltonian is a constant of motion, we have
expressed the energy as a product of two independent constant of
motions. Then, we have seen that these factors are related with
first order ODE's that allow us to get the solutions of the
nonlinear second order ODE. Remark that the Lagrangian underlying
the nonlinear system also permits to get solutions of the system.
There are some interesting papers in the literature, where starting
with the Lagrangian show how to obtain compactons or kink-like
travelling wave solutions of some nonlinear equations
\cite{arodz,adam,gaeta1,gaeta2,gaeta3}.

 \section*{Acknowledgments}
Partial financial support is acknowledged to Junta de Castilla y
Le\'on (Spain) Project GR224. The author acknowledges to Dr. Javier
Negro for useful discussions.

\end{document}